\documentclass[showpacs,preprintnumbers,amsmath,amssymb,prb,twocolumn]{revtex4}

\usepackage{amsfonts}
\usepackage{mathrsfs}
\usepackage{graphicx}
\usepackage{dcolumn}
\usepackage{bm}
\usepackage{exscale}
\usepackage{relsize}
\usepackage{float}

\makeatletter

\newcommand{\Rmnum}[1]{\expandafter\@slowromancap\romannumeral #1@}
\makeatother

\begin{document}
\title{Spin-current Seebeck effect in quantum dot systems }

\author{Zhi-Cheng Yang}
\affiliation{International Center for Quantum Materials, School of Physics, Peking University, Beijing 100871, China }
\author{Qing-Feng Sun}
\email{sunqf@iphy.ac.cn}
\affiliation{International Center for Quantum Materials, School of Physics, Peking University, Beijing 100871, China }
\author{X. C. Xie}
\affiliation{International Center for Quantum Materials, School of Physics, Peking University, Beijing 100871, China }
\date{\today}

\begin{abstract}
  We first bring up the concept of spin-current Seebeck effect based on a recent experiment [Nat. Phys. {\bf 8}, 313 (2012)], and investigate the spin-current Seebeck effect in quantum dot (QD) systems. Our results show that the spin-current Seebeck coefficient $S$ is sensitive to different polarization states of QD, and therefore can be used to detect the polarization state of QD and monitor the transitions between different polarization states of QD. The intradot Coulomb interaction can greatly enhance the $S$ due to the stronger polarization of QD. By using the parameters for a typical QD, we demonstrate that the maximum $S$ can be enhanced by a factor of 80. On the other hand, for a QD whose Coulomb interaction is negligible, we show that one can still obtain a large $S$ by applying an external magnetic field.

\end{abstract}


\maketitle

\section{Introduction}
   The phenomenon where a temperature gradient across a conductor generates an electric voltage is called the Seebeck effect\cite{1}. This effect has found its application in thermal sensing devices such as thermocouple\cite{1,2}. The efficiency of Seebeck effect is measured by the Seebeck coefficient $S$, which is defined as: $V=-S(T_2-T_1)$. Here $V$ is the electric voltage, and $T_2$, $T_1$ are temperatures of the hot and cold regions, respectively. Seebeck coefficient is a powerful tool to characterize materials because it provides information about the energetic difference between the relevant transport states and the Fermi level\cite{3}, while conductance only reflects the density of states near the Fermi level. The Seebeck coefficient $S$ is mainly decided by the energy dependence of the electron scattering in bulk materials. Much effort has been devoted to increasing $S$, for the purpose of enhancing thermoelectric properties\cite{20,21,add1}.

   Recently, Uchida \textit {et al.}\cite{4} has observed a similar effect, where a temperature gradient in a metallic magnet can generate a spin voltage. This effect is known as the spin Seebeck effect, and it allows one to generate a pure spin current without electric currents, which is crucial for spintronic devices\cite{5,6}.
   \begin{figure}[!b]
   \centering
   \includegraphics[width=0.55\textwidth]{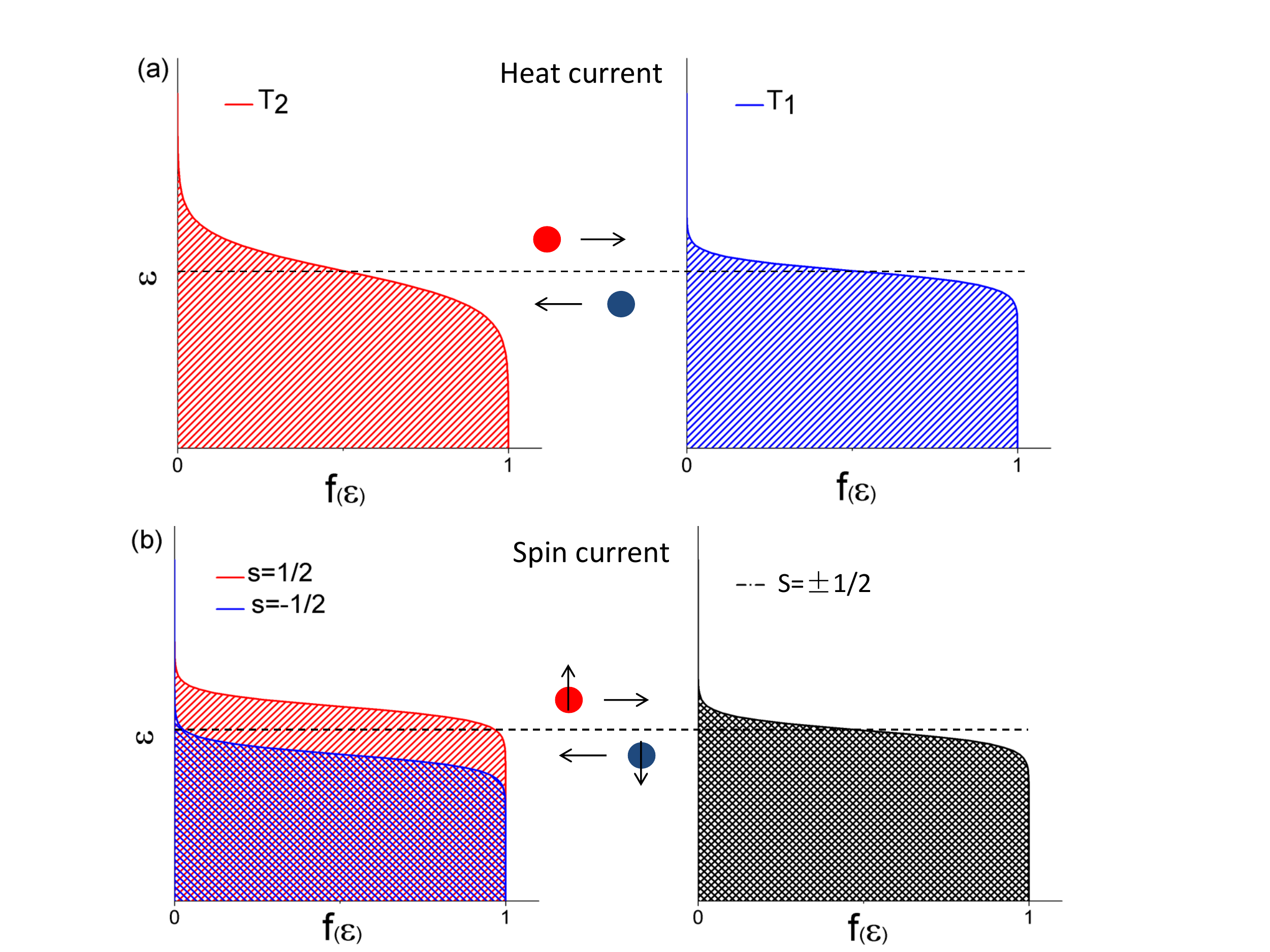}
   \caption{ (Color online) Analogy between Seebeck effect and spin-current Seebeck effect. (a) Seebeck effect; (b) spin-current Seebeck effect.}
   \label{f1}
   \end{figure}

   Here, we investigate another effect, which we call the spin-current Seebeck effect. In analogy to the Seebeck effect, where a heat current generates a charge voltage, here a spin current can also generate a charge voltage\cite{7}. As shown in FIG.\ref{f1}(a), under closed-circuit condition, when a temperature gradient exists, a heat current is set up, with high-energy electrons moving from the left side to the right side and low-energy electrons moving in the opposite direction. When the transmission coefficient is energy dependent, this results in a non-zero net electric current. Under open-circuit condition, an electric voltage is built up for the net electric current to be zero. Now consider spin transport (FIG.\ref{f1}(b)). When a spin bias exists\cite{add3,add4}, a spin current is set up, with spin-up electrons moving from the left side to the right side and spin-down electrons moving in the opposite direction. When the transmission coefficient is energy dependent, the net electric current: $J_e=e(J_{\uparrow}+J_{\downarrow})$ is typically non-zero. Under open-circuit condition, an electric voltage is built up. Note that in both Seebeck effect and spin Seebeck effect, the driving force of the system is heat current, while in spin-current Seebeck effect, the driving force is spin current. On the other hand, both Seebeck effect and spin-current Seebeck effect result in an electric voltage, while the spin Seebeck effect results in a spin voltage. As in the case of Seebeck effect, we can also define a dimensionless spin-current Seebeck coefficient: $V=-(S/e)(\Delta \mu_2-\Delta \mu_1)$, where $\Delta \mu$ is the spin splitting of the chemical potential\cite{7}. Recently, spin-current Seebeck effect has been observed in graphene system\cite{7}, where large values of $\Delta \mu$ can be obtained. Using non-magnetic electrodes, they were able to measure the second-order component of signal: $V \propto I^2$. Experimental results  and numerical modelling are in good agreement, indicating that the signal measured arises owing to the spin-current Seebeck effect.

   Quantum dot (QD) is the simplest and fundamental structure in low-dimensional and mesoscopic transport devices. The energy levels in a QD are well-separated due to the confinement in all three dimensions\cite{8}. A QD is sometimes referred to as a zero-dimensional system and behaves in many ways as an artificial atom\cite{22,23}. The transport properties of a QD can be measured by coupling it to the leads and passing current through it, with the strength of couplings, the number of electrons in the dot, and energy levels in the dot under experimental control\cite{22}. Since 1990s, many novel and interesting transport phenomena in QD systems have been observed in experiments, such as Coulomb blockade\cite{9,10,14,16} and Kondo effect\cite{11,12,13,add2}. QD systems can be described by an Anderson model of a site weakly coupled to ideal leads with an on-site Coulomb interaction $U$.\cite{14} The intradot e-e interaction plays an important role in all the above-mentioned transport phenomena of QD, such as the conductance oscillation at low temperatures in the Coulomb blockade effect\cite{14,16}.

\begin{figure}[!t]
   \centering
   \includegraphics[width=0.45\textwidth]{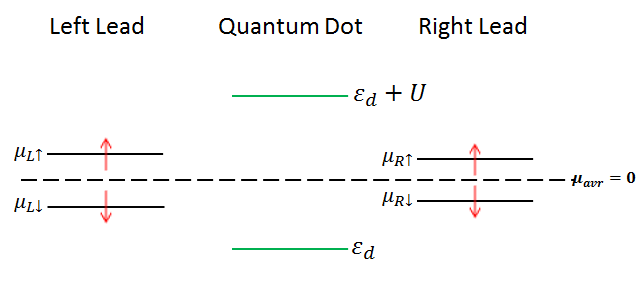}
   \caption{(Color online) Schematic configuration of lead-QD-lead system in this paper.}
   \label{f2}
   \end{figure}

   In this paper, we study the spin-current Seebeck effect of a lead-QD-lead system. The schematic configuration of our system is shown in FIG.\ref{f2}.
   We assume that the spin splitting of chemical potential in the left lead is slightly larger than the right lead, thus setting up a spin bias across the QD. The energy levels of the QD can be tuned by changing the gate voltage $V_g$, thereby changing the energy dependence of the transmission coefficient of the dot. By using the non-equilibrium Green's functions\cite{8,17}, we have obtained an analytical expression of spin-current Seebeck coefficient $S$ in the linear response regime. Applying this expression, we numerically calculated $S$ as a function of $V_g$ under different conditions. The $S-V_g$ curve is always antisymmetric due to the particle-hole symmetry, i.e. the symmetry between electron and hole with opposite spins. Our results demonstrate that the spin-current Seebeck coefficient is very sensitive to the polarization state of the QD, thus one can detect the polarization state of the dot by measuring $S$, while other physical quantities such as conductance and electric current cannot provide information about polarization. The intradot e-e interaction $U$ can greatly enhance $S$ due to the stronger polarization of the dot. For a typical QD where the e-e interaction $U$ is larger by an order than the linewidth $\Gamma$, the $S$ can be enhanced by a factor of 80 comparing with a non-interacting QD. On the other hand, for a QD whose intradot Coulomb interaction is negligible, we show that a large $S$ can still be obtained by applying an external magnetic field.

   \section{Model and Formulation}
    Our lead-QD-lead system in FIG.\ref{f2} can be described by the following Hamiltonian:
     \begin{eqnarray}
      H&=&H_l+H_c+H_t   \nonumber \\
       \nonumber   \\
      H_l&=&\sum_{k,\alpha \in L,R \ \sigma}\epsilon _{k\alpha} c_{k\alpha\sigma}^{\dagger} c_{k\alpha\sigma} \nonumber \\
        \nonumber   \\
      H_c&=&\sum_{\sigma} \epsilon_{d\sigma}d_{\sigma}^{\dagger} d_{\sigma} + Un_{\uparrow}n_{\downarrow}  \nonumber \\
        \nonumber   \\
      H_t&=&\sum_{k,\alpha \in L,R;\sigma}\left [V_{k\alpha} c_{k\alpha\sigma}^{\dagger}d_{\sigma} +h.c.\right]
      \label{eq1}
      \end{eqnarray}
    Here $H_l$ describes the non-interacting leads, $\alpha=L,R$ represents the left and right leads respectively. $c^{\dagger}_{k\alpha \sigma}$($c_{k\alpha \sigma}$) creates (annihilates) an electron with spin $\sigma$ in the $\alpha$ lead. $H_c$ describes the QD with one single-particle energy level $\epsilon_{d\sigma}$ and intradot Coulomb interaction $U$. The single-particle energy level $\epsilon_{d\sigma}$ is spin-degenerate in the absence of an external magnetic field $B$. When applying an external magnetic field $B$, $\epsilon_{d\sigma}=\epsilon_d - \sigma B$ due to the Zeeman splitting. $H_t$ is the Hamiltonian for tunneling between the leads and QD, where $V_{k\alpha}$ is the tunneling matrix element. Note that the notation we use here means we have assumed there is no spin-flipping mechanism in our system, so the spin of an electron does not change when tunneling between the leads and QD.

   By using non-equilibrium Green's function, the electric current with spin $\sigma$ polarization can be written as\cite{17,18}:
    \begin{equation}
     J_{\sigma}=\frac{ie}{\hbar}\int \frac{d\epsilon}{2\pi} \left [f_{L\sigma}(\epsilon)-f_{R\sigma}(\epsilon) \right ] \mathcal{T}_{\sigma}(\epsilon)
    \label{eq2}
    \end{equation}
where $f_{\alpha \sigma}(\epsilon)=1/\{exp[(\epsilon-\mu_{\alpha \sigma})/k_BT]+1\}$ is the Fermi distribution function of the spin $\sigma$ electrons in the $\alpha$ lead. $\mathcal{T}_{\sigma}(\epsilon)$ is the transmission coefficient for spin $\sigma$ electrons. $\mathcal{T}_{\sigma}(\epsilon)$ can be written as:
     \begin{equation}
     \mathcal{T}_{\sigma}(\epsilon)=Tr \left \{ \frac {\Gamma^L(\epsilon)\Gamma^R(\epsilon)}{\Gamma^L(\epsilon)+\Gamma^R(\epsilon)}\left[G^r_{\sigma}(\epsilon)-G^a_{\sigma}(\epsilon)\right]\right \}
     \end{equation}
    Here $G^r(\epsilon)\left(G^a(\epsilon)\right)$ is the standard retarded (advanced) Green's function of the QD in the presence of coupling to the leads. $\Gamma^{\alpha}(\epsilon)=2\pi \sum_k |V_{k\alpha}|^2\delta (\epsilon-\epsilon_{k\alpha})$ are the linewidth functions. The Green's functions can be obtained from the Dyson equation and Keldysh formalism\cite{17,19}:
      \begin{eqnarray}
     {\bf G}^r(\epsilon) \equiv \
     \left (
     \begin{array}{cc}

     G_{\uparrow}^r(\epsilon)  &  0 \\
       0     &  G_{\downarrow}^r(\epsilon)\\
    \end{array}
    \right )
={\bf g}^r(\epsilon)+{\bf g}^r(\epsilon){\bf \Sigma}^r{\bf G}^r(\epsilon)
     \label{eq4}   \\
     {\bf G}^<(\epsilon) \equiv
     \left (
     \begin{array}{cc}

     G_{\uparrow}^<(\epsilon)  &  0 \\
       0     &  G_{\downarrow}^<(\epsilon)\\

    \end{array}
    \right )
   ={\bf G}^r(\epsilon){\bf \Sigma}^<(\epsilon){\bf G}^a(\epsilon)
    \end{eqnarray}
  Here the boldface letters $({\bf G}$, ${\bf g}$, and ${\bf \Sigma}$) represent the $2 \times 2$ matrices. ${\bf g}^r$ is the Green's function of QD without coupling to the leads. ${\bf \Sigma}^{r,<}$ are self-energies (we have neglected the higher order of self-energy correction that originates from the combination of the e-e interaction and the tunneling terms.)\cite{19}:
     \begin{eqnarray}
    &&{\bf \Sigma}^r=
    \left (
    \begin{array}{cc}
     i\Gamma & \ 0 \\
      0  & \  i\Gamma \\
     \end{array}
    \right )     \\
    &&{\bf \Sigma}^<=
    \left (
    \begin{array}{cc}
    i\Gamma(f_{L\uparrow}+f_{R\uparrow})  &  0 \\
     0  &  i\Gamma(f_{L\downarrow}+f_{R\downarrow})\\
    \end{array}
    \right )
    \end{eqnarray}
  where we have neglected the energy dependence of linewidth functions $\Gamma^{\alpha}$ and consider symmetric barriers: $\Gamma^L=\Gamma^R=\Gamma$. ${\bf g}^r$ can be calculated applying the equation-of-motion technique, and hence ${\bf G}^r$ can be obtained from Eq.(\ref{eq4}):
   \begin{eqnarray}
     g^r_{\sigma}(\epsilon)&=&\frac {\langle n_{\bar{\sigma}}\rangle}{\epsilon-\epsilon_d-U}+\frac{1-\langle n_{\bar{\sigma}}\rangle}{\epsilon-\epsilon_d}   \\
           \nonumber \\
     G^r_{\sigma}(\epsilon)&=&\frac{g^r_{\sigma}(\epsilon)}{1-g^r_{\sigma}(\epsilon)\Sigma^r_{\sigma}} \nonumber \\
        \nonumber \\
&=&\frac {\epsilon-\epsilon_d-(1-\langle n_{\bar{\sigma}}\rangle)U}{(\epsilon-\epsilon_d-U)(\epsilon-\epsilon_d)-\Sigma^r_{\sigma}[\epsilon-\epsilon_d-(1-\langle n_{\bar{\sigma}}\rangle)U]} \nonumber \\
         \nonumber \\
&\approx&  \frac{1-\langle n_{\bar{\sigma}}\rangle}{\epsilon-\epsilon_d-\Sigma^r_{\sigma}}+\frac{\langle n_{\bar{\sigma}}\rangle}{\epsilon-\epsilon_d-U-\Sigma^r_{\sigma}}
     \label{eq9}
     \end{eqnarray}
 where $\langle n_{\bar {\sigma}} \rangle$ is the occupation number of electrons with spin ${\bar {\sigma}}$, ${\bar {\sigma}}=\downarrow$ while $\sigma=\uparrow$ and ${\bar {\sigma}}=\uparrow$ while $\sigma=\downarrow$. $\langle n_{\sigma}\rangle$ is related to the Green's function via: $\langle n_{\sigma} \rangle=\int \frac{d\epsilon}{2\pi i} G^<_{\sigma}(\epsilon)$, thus Eq.(\ref{eq4})-Eq.(\ref{eq9}) need to be solved self-consistently.

   Now we consider the spin-current Seebeck coefficient of our system in FIG.\ref{f2}. $\mu_{avr}$ can be conveniently chosen to be zero, thus we have $\Delta \mu_{2}=\mu_{L\uparrow}=-\mu_{L\downarrow}$, $\Delta \mu_{1}=\mu_{R\uparrow}=-\mu_{R\downarrow}$. We define spin bias as: $\Delta V_s=\Delta \mu_{2}-\Delta \mu_{1}=\mu_{L\uparrow}-\mu_{R\uparrow}=\mu_{R\downarrow}-\mu_{L\downarrow}$. As discussed in Part \Rmnum{1}, $\Delta V_s$ will induce an electric voltage $V$ under open-circuit conditions. If we consider linear response regime, i.e. $\Delta V_s \rightarrow 0$, we can keep only the first-order terms of Eq.(\ref{eq2}), which yields the following expression for the spin-current Seebeck coefficient $S$:
     \begin{equation}
     S=-\frac{\mathlarger \int d\omega \left[ \left (\frac {\partial f}{\partial \omega}\right )_{\mu=\mu_{R\uparrow}}
\mathcal{T}_{\uparrow}(\omega)-\left (\frac{\partial f}{\partial \omega} \right )_{\mu=\mu_{R\downarrow}}\mathcal{T}_{\downarrow}(\omega) \right ]} {\mathlarger \int d\omega \left[ \left (\frac {\partial f}{\partial \omega}\right )_{\mu=\mu_{R\uparrow}}
\mathcal{T}_{\uparrow}(\omega)+\left (\frac{\partial f}{\partial \omega} \right )_{\mu=\mu_{R\downarrow}}\mathcal{T}_{\downarrow}(\omega) \right ]}
    \label{eq10}
    \end{equation}
   This formula plays the role as the starting point for the following numerical calculations.

   \section{Numerical Results}
   In numerical investigation, we consider symmetric barriers: $\Gamma^L=\Gamma^R=\Gamma$ as mentioned above, and set $\Gamma=1$ and $\mu_{avr}=0$ as the energy reference. In the linear regime, we can define $V_s \equiv \mu_{L\uparrow}=\mu_{R\uparrow}=-\mu_{L\downarrow}=-\mu_{R\downarrow}$. Note that linear regime means $\Delta V_s \rightarrow 0$, but $V_s$ does not necessarily tend to zero. Below, we also study non-interacting QD, i.e. $U=0$. In this case, the Green's function of QD in the presence of coupling can be exactly obtained as $G^r_{\sigma}(\epsilon)=G^{a*}_{\sigma}(\epsilon)=1/(\epsilon-\epsilon_{d\sigma}+i\Gamma)$, thus the tunneling coefficient and spin-current Seebeck coefficient can be calculated straightforwardly. We first study the QD in the absence of external magnetic field, hence the single-electron energy level is spin-degenerate.
   \begin{figure}[!h]
   \centering
   \includegraphics[width=0.5\textwidth]{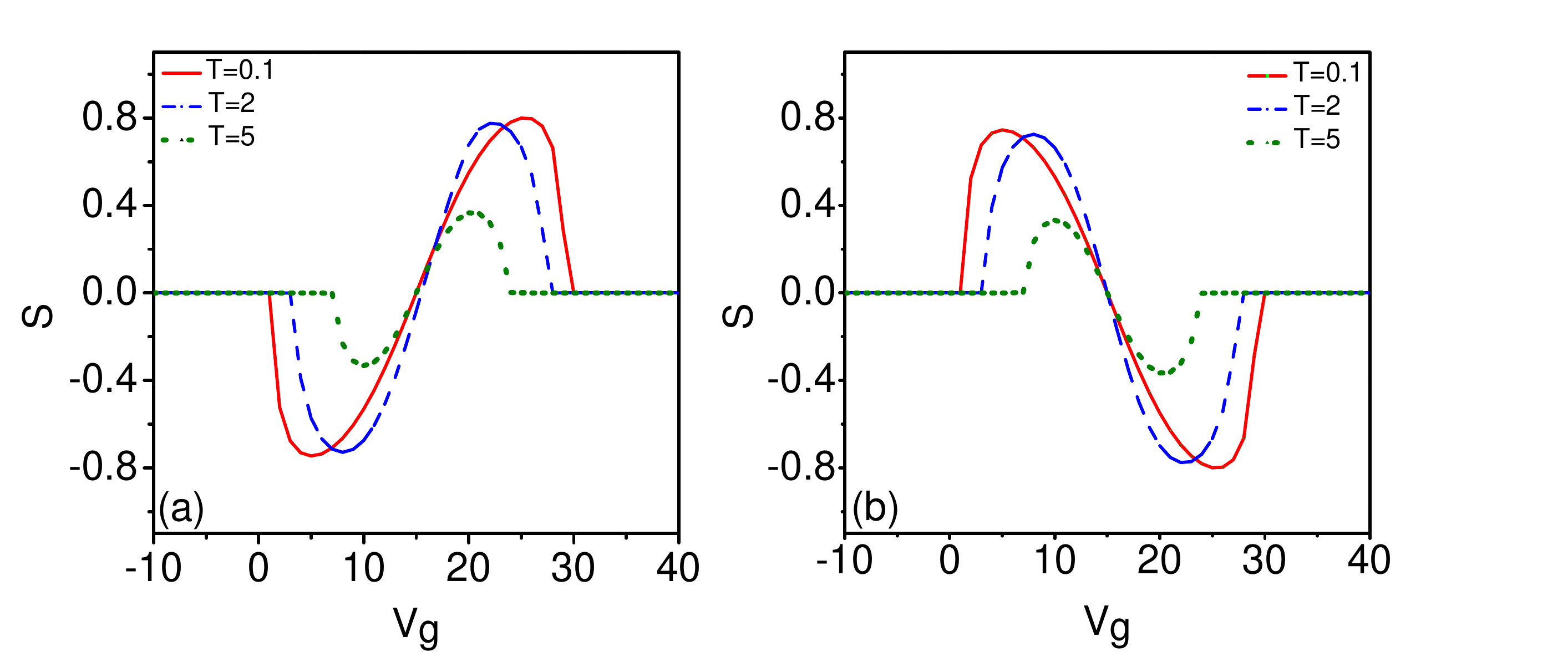}
   \caption{ (Color online) $S$ vs $V_g$ for different polarization states of QD at different temperatures with $V_s=0$ and $U=30$. (a) $\langle n_{\uparrow}\rangle>\langle n_{\downarrow} \rangle$; (b) $\langle n_{\uparrow}\rangle < \langle n_{\downarrow} \rangle$.}
   \label{f3}
   \end{figure}

   We start our calculation from the case where $V_s=0$. It is worthy to note that in the traditional Seebeck effect, the temperature $T$ can never actually reach zero. Here, in the spin-current Seebeck effect, there is no problem with $V_s$ being zero. When $U=0$, the spin-up and spin-down electrons are completely symmetric in the system, hence $S\equiv 0$. When $U$ is large enough, QD will become spin-polarized, then $S$ can emerge. Since the lead-QD-lead system can be modeled by an Anderson-type Hamiltonian, we learn that Eq.(\ref{eq4})-Eq.(\ref{eq9}) can yield three sets of solutions for $\langle n_{\uparrow}\rangle, \langle n_{\downarrow} \rangle$: i. $\langle n_{\uparrow}\rangle=\langle n_{\downarrow} \rangle$; ii. $\langle n_{\uparrow}\rangle>\langle n_{\downarrow} \rangle$; iii. $\langle n_{\uparrow}\rangle<\langle n_{\downarrow} \rangle$. Solution i. represents a high-energy state of the system with QD unpolarized, so $S=0$. Solution ii. and iii. are symmetric with opposite spin-polarization states of QD, as shown in FIG. \ref{f3}. It can be seen that the $S-V_g$ curves corresponding to the two different polarization states of QD are opposite. This important feature indicates that: (1) $S$ is sensitive to the polarization state of QD, and can be used to distinguish between state ii. and state iii., while other quantities such as electric current and conductance cannot tell the difference ; (2) When a transition from state ii. to state iii. occurs, $S$ will change sign; thus one can monitor the transitions between the two states by measuring $S$.

\begin{figure}[!b]
   \centering
   \includegraphics[width=0.5\textwidth]{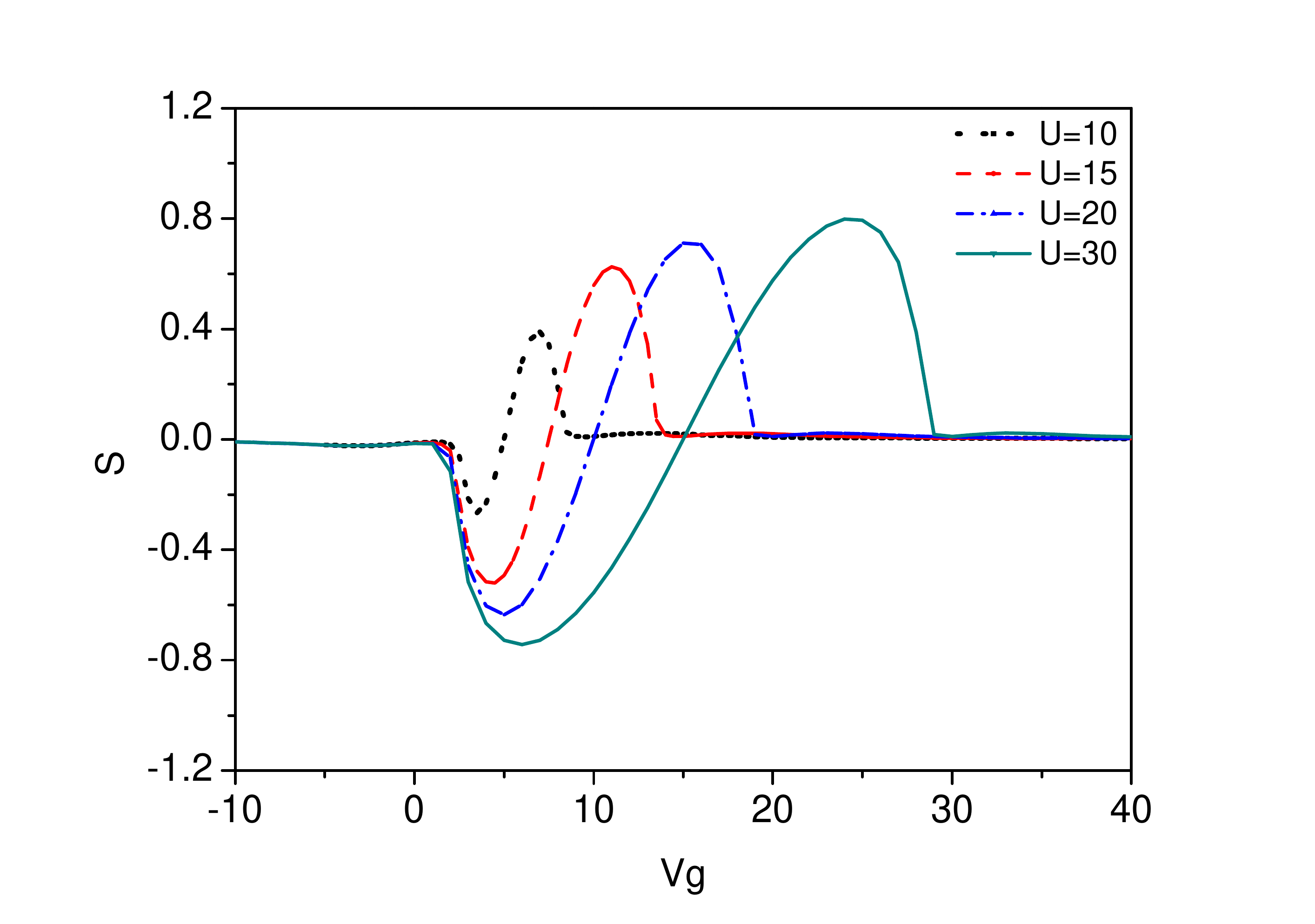}
   \caption{ (Color online) $S$ vs $V_g$ at different $U$. $V_s=0.04$, $T=1$.}
   \label{f5}
   \end{figure}

   Next we consider the case where $V_s$ is small. It can be readily learned from Anderson model that the system has three sets of solutions when $V_s=0$. We further show via numerical approach that the system still have three sets of solutions when $V_s$ is non-zero:  i. $\langle n_{\uparrow}\rangle \lesssim \langle n_{\downarrow} \rangle$; ii. $\langle n_{\uparrow}\rangle>\langle n_{\downarrow} \rangle$; iii. $\langle n_{\uparrow}\rangle<\langle n_{\downarrow} \rangle$. Hereafter we choose the solution: $\langle n_{\uparrow}\rangle>\langle n_{\downarrow} \rangle$ for definiteness.
    \begin{figure}[!t]
   \centering
   \includegraphics[width=0.5\textwidth]{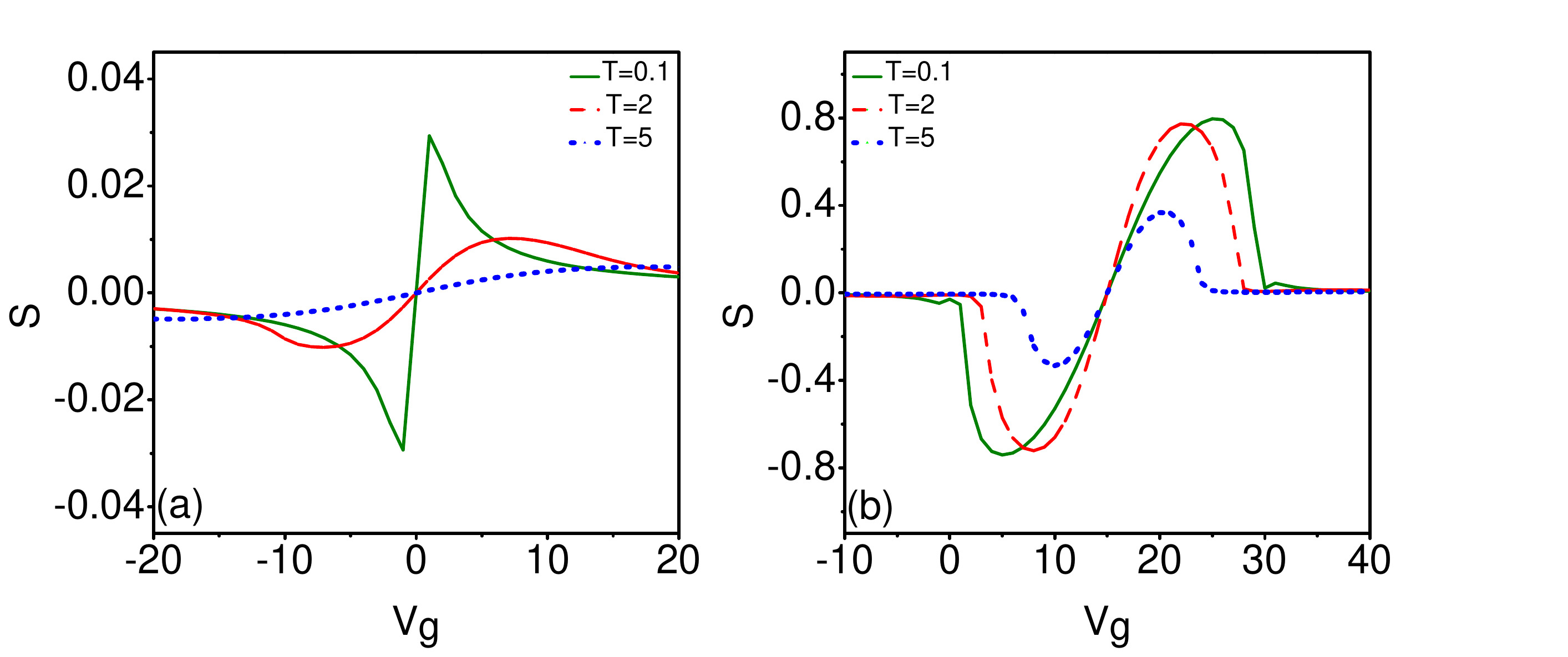}
   \caption{ (Color online) $S$ vs $V_g$ at different temperatures. $V_s=0.04$. (a) $U=0$; (b) $U=30$.}
   \label{f4}
   \end{figure}
   FIG. \ref{f4} shows the $S$ versus $V_g$ at different temperatures for $U=0$ and $U=30$, respectively. We find that $S$ is greatly enhanced when $U$ is large, by a maximum factor of 80. Also, $S$ is more robust against temperature increase at larger $U$. The reason is as follows. When $U=0$, the QD is unpolarized, and the density of state (proportional to the transmission coefficient) is identical for spin-up and spin-down electrons in the QD. When $U\neq 0$, the transmission coefficient for electrons with different spins near the Fermi level will become rather different because of the spin polarization of QD, and the difference will become more distinct as $U$ increases. To get a clearer view of the effect of $U$, we show the $S$ versus $V_g$ at different $U$ in FIG. \ref{f5},
   where we can see that $S$ is enhanced as $U$ increases. Moreover, the region with apparently non-zero $S$ corresponds to the Coulomb blockade regime, with occupation number of QD $\langle n_{\uparrow} \rangle + \langle n_{\downarrow} \rangle \approx 1$; while the left and right regions with nearly zero $S$ corresponds to an occupation number of 0 and 2 (the empty state and double-occupied state) respectively.

   We further notice that $S$ is always antisymmetric under various values of all parameters. This is due to the particle-hole symmetry of our system. More specifically, one can show that the Hamiltonian (\ref{eq1}) is invariant under the following particle-hole transform: $d_{\sigma} \rightarrow \tilde{d}^{\dagger}_{\bar {\sigma}}$, $d_{\sigma}^{\dagger} \rightarrow \tilde{d}_{\bar {\sigma}}$, provided that $2\epsilon_d + U = 0$, which is exactly the symmetric center of $S-V_g$ curves shown above. Also, the Fermi distribution function of the leads: $f(\epsilon-\mu_{\sigma}) \rightarrow f(-\epsilon+\mu_{\bar {\sigma}})=1-f(\epsilon-\mu_{\bar {\sigma}})$. So the electrons and holes with opposite spins are symmetric in our system, which yields the antisymmetric behavior of $S$.

\begin{figure}[!h]
   \centering
   \includegraphics[width=0.5\textwidth]{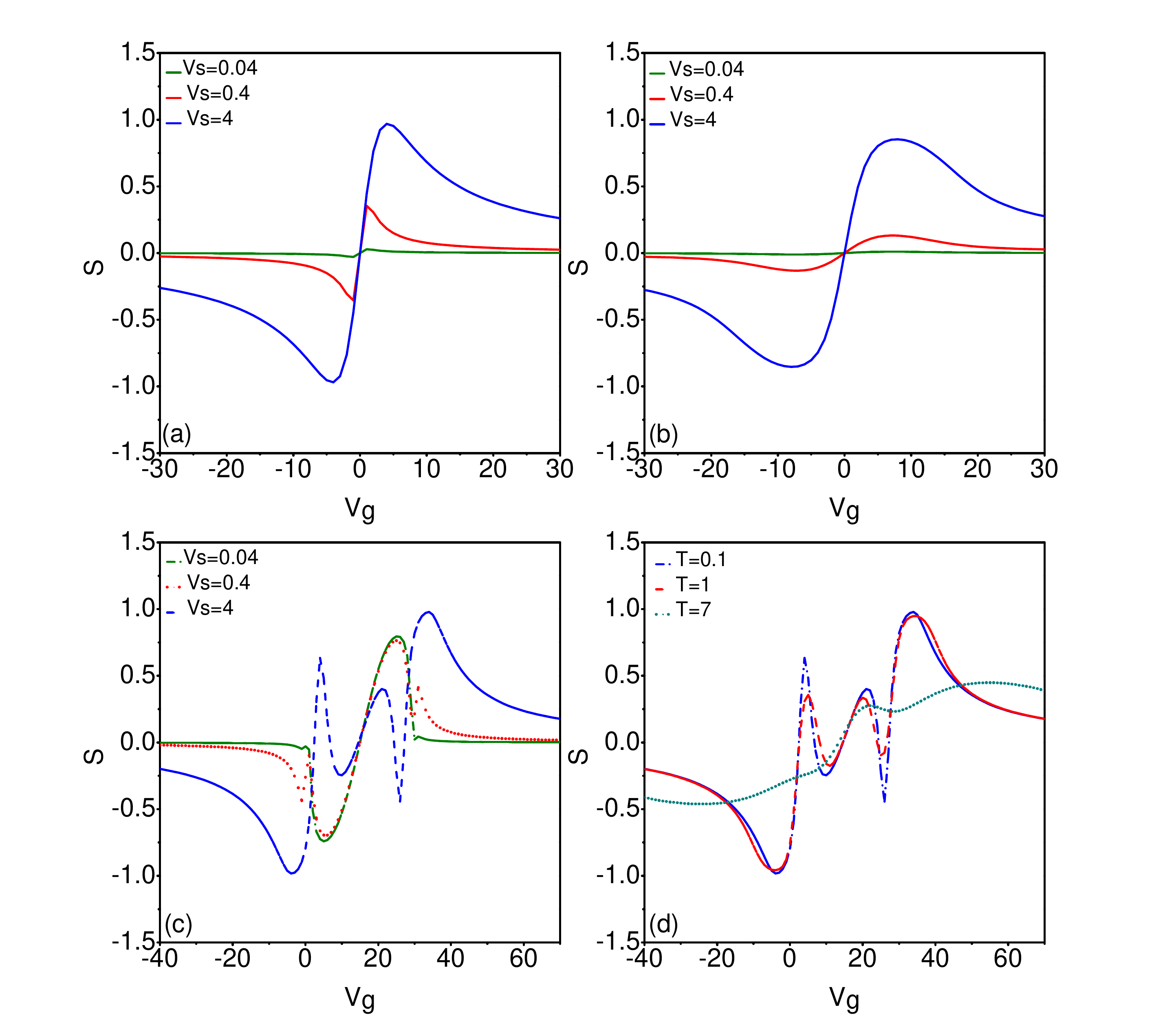}
   \caption{ (Color online) $S$ vs $V_g$ at different temperatures and $V_s$. For (a) and (b), $U=0$; for (c) and (d), $U=30$; (a) $T=0.1$; (b) $T=2$; (c) $T=0.1$; (d) $V_s=4$; }
   \label{f6}
   \end{figure}

  As we have emphasized at the beginning of this part, although we take the limit $\Delta V_s \rightarrow 0$ in linear regime, $V_s$ does not have to be small. FIG. \ref{f6} takes into account the case where $V_s$ is large.
   FIG. \ref{f6}(a) and FIG. \ref{f6}(b) depicts the $S$ versus $V_g$ at different $V_s$, with the temperature $T=0.1$ and $T=2$ respectively. We find that $S$ is greatly enhanced when $V_s$ is enlarged; in addition, when $T$ is lower, $S$ begins to increase at smaller $V_s$. The reasons can be explained as follows. When $V_s$ is enlarged, the Fermi levels for different spins are well separated, which leads to a distinct difference in transmission coefficient for spin-up and spin-down electrons, hence $S$ is enhanced. The excited states of electrons above Fermi energy due to the increase of temperature will offset the separation of Fermi levels, so $S$ begins to increase at smaller $V_s$ at lower temperatures. It is interesting to note that, by comparing FIG. \ref{f6}(a) with FIG. \ref{f6}(c), new peaks begin to emerge at large $V_s$ in the existence of $U$. These new peaks emerge when one of the energy levels in QD coincides with either Fermi level in the lead, while the other energy level in QD as well as the other Fermi level is far away, so new peaks emerge only when both $V_s$ and $U$ are large. FIG. {\ref{f6}(d) indicates that thermal fluctuations caused by increase of temperature will smear out all structures of peaks.

   So far, we have been studying QD in the absence of external magnetic field, therefore the single-particle energy level in the dot is spin-degenerate. When applying an external magnetic field $B$, $\epsilon_{d\sigma}=\epsilon_d - \sigma B$ due to the Zeeman splitting. For simplicity, we consider only non-interacting QD and set $U=0$. FIG. \ref{f7} shows $S$ at different $B$, with $V_s=0$.
   \begin{figure}[!t]
   \centering
   \includegraphics[width=0.5\textwidth]{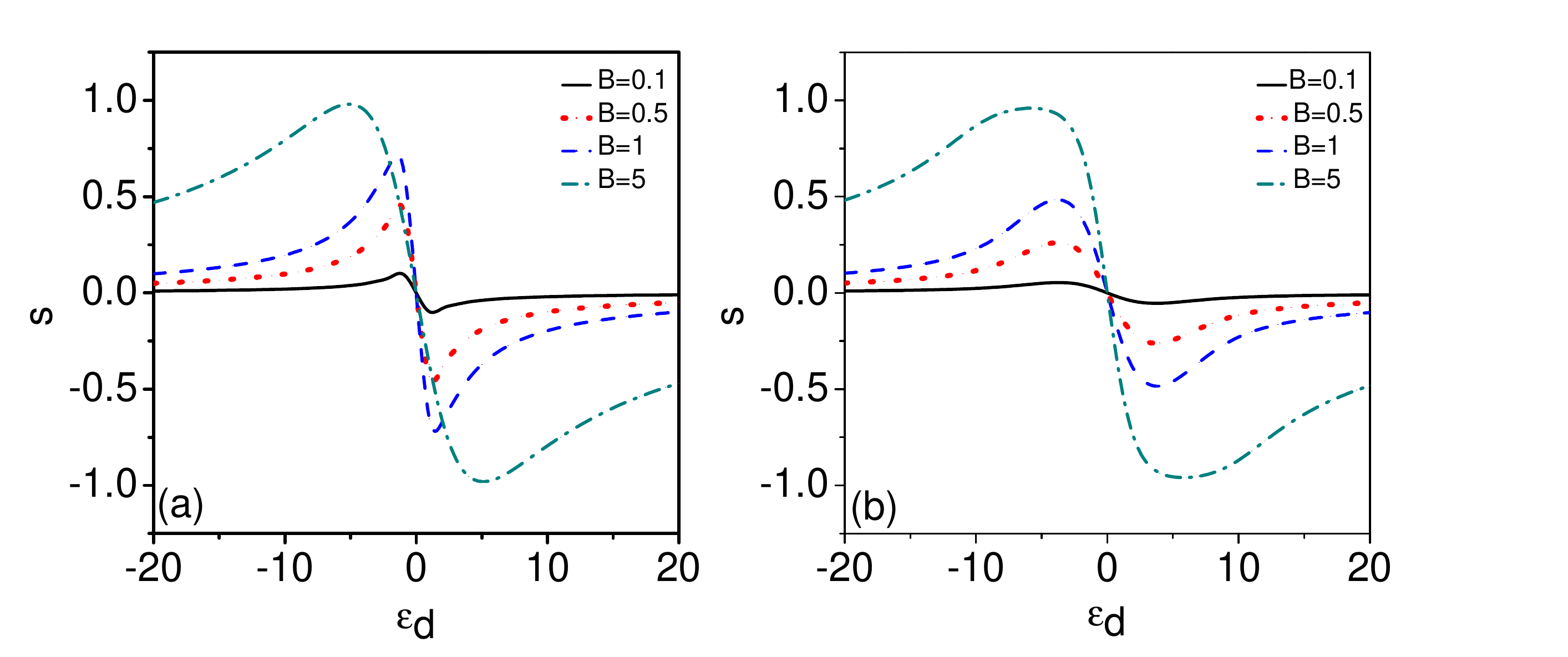}
   \caption{ (Color online) $S$ vs $\epsilon_d$ at different $B$ with  $V_s=0$ and $U=0$. (a) $T=0.1$; (b) $T=1$. }
   \label{f7}
   \end{figure}
   From the discussions above, we learn that in the case of non-interacting QD, $S\equiv 0$ when $V_s=0$ and $B=0$. Here we find that $S$ is greatly enhanced upon applying a magnetic field, even for a non-interacting QD at $V_s=0$. This indicates that we are still able to achieve large $S$ for QD whose intradot e-e interactions are negligible by applying an external magnetic field. In addition, it is worthy to note that, comparing to FIG. \ref{f3}, $S$ is obviously non-zero here even in the regions with an occupation number of 0 and 2. This can be understood as follows. In FIG. \ref{f3} the tunneling coefficient for each spin has two resonant peaks, located at $\epsilon_d$ and $\epsilon_d+U$ respectively; while in the case here the tunneling coefficient for each spin has only one resonant peak. Therefore, when the Fermi energy level is close to one resonant peak, the transmission coefficient for the other spin is much lesser, resulting in a non-zero $S$ in these regions.

   Since the results for small $V_s$ is similar to $V_s=0$, we hereby study the case when $V_s$ is large, as shown in FIG. \ref{f8}. We find that no new peak emerges at large $V_s$ comparing with FIG. \ref{f6}(c), because there is only one resonant peak for the transmission coefficient for each spin. Also, $S$ is more robust against temperature in the existence of magnetic field.
   \begin{figure}[H]
   \centering
   \includegraphics[width=0.5\textwidth]{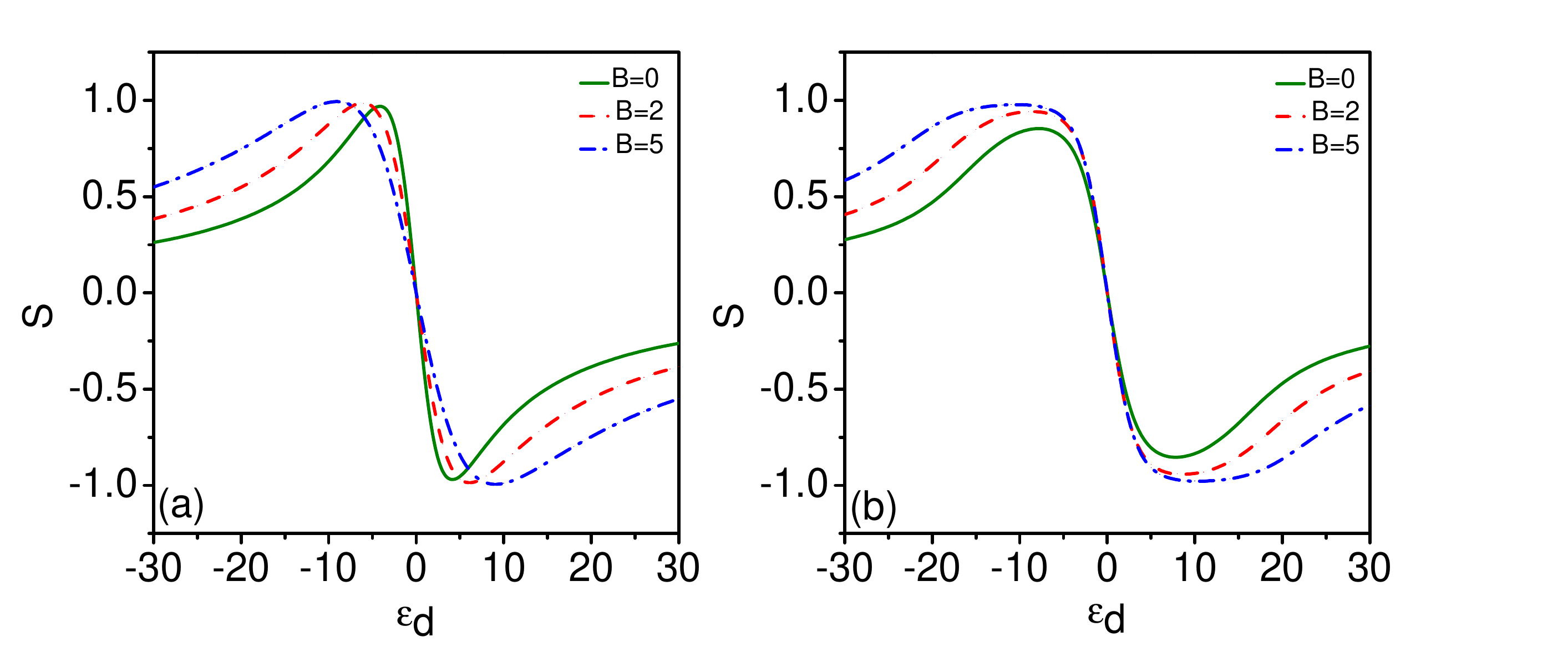}
   \caption{ (Color online) $S$ vs $\epsilon_d$ at different $B$ with  $V_s=4$ and $U=0$. (a) $T=0.1$; (b) $T=2$. }
   \label{f8}
   \end{figure}

    \section{Conclusions}
    We investigate the spin-current Seebeck effect in quantum dot systems. We show that the spin-current Seebeck coefficient $S$ is always antisymmetric, originating from the particle-hole symmetry of our system. Our results demonstrate that $S$ is sensitive to different polarization states of QD, thus can be used to distinguish between different states and monitor the transitions between them. The intradot e-e Coulomb interaction $U$ can greatly enhance $S$ due to the stronger polarization of QD induced by $U$. For a typical QD whose $U$ is larger by an order than $\Gamma$, the maximum $S$ can be enhanced by a factor of 80. For a QD whose intradot Coulomb interaction is negligible, our work demonstrates that a large $S$ can still be obtained by applying an external magnetic field.

     \section*{ACKNOWLEDGMENTS}
    Z.-C. Yang is deeply indebted to Dr. Haiwen Liu and Dr. Hua Jiang for inspiring dicusssions. This work was financially supported by NBRP of China (2012CB921303, 2009CB929100 and 2012CB821402), NSF-China, under Grants Nos. 11074174, 11121063 and 11274364.

\end{document}